\documentclass[osajnl,twocolumn,showpacs,superscriptaddress,10pt]{revtex4-1} 
\usepackage{amsmath,amssymb,graphicx}

\begin{document}

\title{Strong Raman-induced non-instantaneous soliton interactions in gas-filled photonic crystal fibers}

\author{Mohammed F. Saleh}
\email{Corresponding author: m.saleh@hw.ac.uk}
\affiliation{School of Engineering and Physical Sciences, Heriot-Watt University, EH14 4AS Edinburgh, UK}
\affiliation{Max Planck Institute for the Science of Light, G\"{u}nther-Scharowsky str. 1, 91058 Erlangen, Germany}

\author{Andrea Armaroli}
\affiliation{Max Planck Institute for the Science of Light, G\"{u}nther-Scharowsky str. 1, 91058 Erlangen, Germany}
\affiliation{Laboratoire FOTON, CNRS UMR6082,6, rue de Kerampont, CS80518 22305 Lannion CEDEX France}
\author{Andrea Marini}
\affiliation{Max Planck Institute for the Science of Light, G\"{u}nther-Scharowsky str. 1, 91058 Erlangen, Germany}
\affiliation{ICFO-Institut de Ciencies Fotoniques, Mediterranean Technology Park, 08860 Castelldefels (Barcelona), Spain}

\author{Fabio Biancalana}
\affiliation{School of Engineering and Physical Sciences, Heriot-Watt University, EH14 4AS Edinburgh, UK}
\affiliation{Max Planck Institute for the Science of Light, G\"{u}nther-Scharowsky str. 1, 91058 Erlangen, Germany}

\begin{abstract}
We have developed an analytical model based on the perturbation theory in order to study the optical propagation of two successive intense solitons in hollow-core photonic crystal fibers filled with Raman-active gases. Based on the time delay between the two solitons, we have found that the trailing soliton dynamics can experience unusual nonlinear phenomena such as spectral and temporal soliton oscillations and transport towards the leading soliton. The overall dynamics can lead to a spatiotemporal modulation of the refractive index with a uniform temporal period and a uniform or chirped spatial period. 
\end{abstract}


\maketitle

Raman scattering involves the transfer of a small fraction of energy of an optical field to another field at longer wavelengths via inelastic scattering with optical phonons supplied by the medium. The Raman self-frequency redshift, first observed in conventional optical fibers  \cite{Dianov85,Mitschke86,Bulushev91}, can continuously downshift the central frequency of a single pulse during propagation. With the invention of solid-silica-core photonic crystal fibers (PCFs) \cite{Russell03,Russell06}, this process has been demonstrated and extensively exploited in attaining a broad supercontinuum \cite{Dudley06}. The main ingredients are the high confinement of light and tunability dispersion provided by these fibers. Gas-filled hollow-core (HC) PCFs with Kagome lattice have become a strong competitor for solid-silica-core PCFs \cite{Travers11,Russell14}. Besides the latter benefits, a wide range of gases with different properties offers several opportunities to demonstrate new phenomena and applications using optical fibers \cite{Benabid02a,Heckl09,Joly11,Chang11,Hoelzer11b,Saleh11a,Saleh12,Chang13,Saleh13,Saleh14,Belli15,Saleh15a}. 

Raman scattering processes in gases are characterized by having very long molecular coherence relaxation (dephasing) time ($ \sim $ps), at least three orders of magnitude higher than silica glass. Within this time, the medium possesses a high \textit{non-instantaneous} nonlinear response. A Raman-induced continuous downshift of the frequency of an ultrashort pulse propagating in a gas has been observed \cite{Korn98}. Moreover, a sinusoidal temporal modulation of the medium refractive index lagging the pulse has been detected by launching a delayed weak probe within the dephasing time \cite{Korn98,Nazarkin99,Wittmann00}. Very recently, we have treated this modulation as a periodic temporal crystal \cite{Saleh15a} and we have proposed to use hydrogen-filled HC-PCFs to observe the analogue condensed matter physics phenomena, such as the Wannier-Stark ladder \cite{Wannier60}, the Bloch oscillations \cite{Bloch28} and Zener tunneling \cite{Zener34}. In this Letter, we investigate the case when the delayed pulse is another strong fundamental soliton rather than a weak probe pulse as in \cite{Saleh15a}. The delayed soliton can be treated as a particle in a moving periodic potential induced by the leading soliton. Phenomena such as spectral and temporal soliton oscillations as well as upgrading the temporal crystal to a spatiotemporal one are predicted here.

Pulse propagation in Raman-active gases is governed by the interplay between the Maxwell equations and the Bloch equations of an effective two-level system describing a single Raman transition. We have previously shown that this set of equations can be simplified to a single generalized nonlinear Schr\"{o}dinger equation for femtosecond pulses with an energy of a few micro-Joules \cite{Saleh15a},
\begin{equation}
i\partial_{\xi}\psi+\frac{1}{2}\partial_{\tau}^{2}\psi+|\psi|^{2}\psi+R\left(\tau \right)\psi=0,
\label{eq1}
\end{equation}
where $ R\left(\tau \right)=\kappa \int_{-\infty}^{\tau}\sin\left[ \delta\left( \tau-\tau'\right) \right]  \left|\psi\left( \tau'\right) \right|^{2} d\tau'$ is the nonlinear Raman convolution, $ \psi $ is the pulse complex envelope, $ \xi $ is the longitudinal coordinate along the fiber, $ \tau $ is the time variable in a reference frame moving with the pulse group velocity, $ \delta=2\pi/\bar{\Lambda} $ is the Raman frequency shift, $ \bar{\Lambda} $  is the corresponding oscillation period, $ \kappa $ is the relative strength between the Raman and Kerr nonlinearities, and the equation is in normalized units. We have assumed propagation in the deep anomalous dispersion regime  to determine the pure effect of the temporal nonlocality of the Raman nonlinearity on successive intense solitons, which can occur during the soliton fission process in supercontinuum generation in gases \cite{Belli15}.

We are interested in investigating the propagation of two successive strong ultrashort pulses, $ \psi=\psi_{1}+\psi_{2}$. The two pulses are assumed to have the same frequency, hence, they will propagate with the same group velocity, and experience the same dispersion. Also, they are not temporally overlapped, but separated by a delay less than the Raman dephasing (relaxation) times. In this case, all the nonlinear cross terms are canceled out, with the exception of the Raman polarization effect of the leading pulse on the trailing pulse. Under these conditions Eq. (\ref{eq1}) can be split into two equations for $ \psi_{1} $ and $ \psi_{2} $, 
\begin{equation}
i\partial_{\xi}\psi_{j}+\frac{1}{2}\partial_{\tau}^{2}\psi_{j}+|\psi_{j}|^{2}\psi_{j}+R_{j}\left(\tau \right)\psi_{j}=0, 
\label{eq2}
\end{equation}
with $ j=1,2 $. For weak Raman nonlinearities, the solutions of these equations are two perturbed fundamental solitons,  $ \psi_{j}\left( \xi,\tau\right)=V_{j}\,\mathrm{sech} \left[V_{j} \left(\tau-\bar{\tau}_{j}\left(\xi \right) \right) \right] \exp\left[-i\Omega_{j}\left(\xi \right)\tau\right] $ where  $V_{j} $, $ \Omega_{j}$, and $\bar{\tau}_{j} $  are the $ j^{\mathrm{th}} $ soliton amplitude, central frequency, and temporal peak, respectively.  When the soliton durations $ \ll 1/\delta $, their Raman response functions can be approximated by using a Taylor expansion \cite{Saleh15a},
\begin{equation}
R_{j}\left(\tau \right)\approx \kappa\sum_{l} V_{l} \sin\left[ \delta\left(\tau-\bar{\tau}_{l} \right)  \right] \left\lbrace  1+ \mathrm{tanh} \left[  V_{l} \left(\tau-\bar{\tau}_{l} \right) \right]  \right\rbrace,
\end{equation}
with $ l=1,2 $. Hence, the leading soliton will induce a retarded sinusoidal Raman polarization that will impact the dynamics of the trailing soliton. 

\begin{figure}
\centerline{\includegraphics[width=8.6cm]{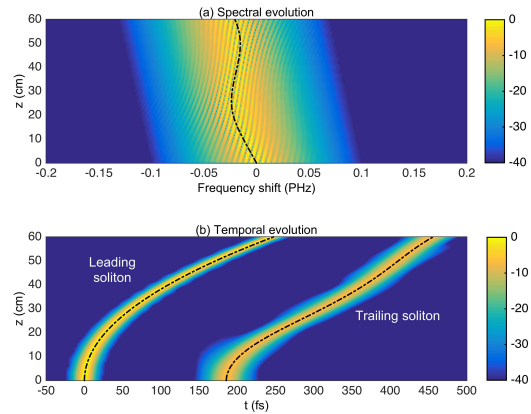}}
\caption{(Color online). (a,b) Spectral and temporal evolution of two fundamental solitons launched in a HC-PCF filled by a Raman-active gas, respectively, with $ \kappa=0.16 $, and $ \delta=0.5 $ (equivalent to  a Raman-mode oscillation with period 185 fs, such as in deuterium \cite{Burzo07}). All subsequent simulations in this Letter use these parameters. The leading and trailing solitons have normalized amplitudes $ V_{1} = 2.5 $, $ V_{2}=  1.25$, and full width half maximum (FWHM) 8 fs, 16 fs, respectively. The black dashed-dotted lines are the analytical predictions using Eq. (\ref{eq3}). Contour plots are given in a logarithmic scale and truncated at -40 dB.
\label{Fig1}}
\end{figure}

To understand how Raman nonlinearities can affect the pulse dynamics, we have adopted the variational perturbation method \cite{Agrawal07}. Using this method, we have considered each Raman contribution $ R_{j} $ as a perturbation for the associated solution (a fundamental soliton $ j $) of its  equation. A set of coupled governing equations that determine the evolution of each soliton parameters can then be derived,
\begin{equation}
\begin{array}{ll}
\Omega_{1} &= - g_{1}\, \xi,\\ 
\bar{\tau}_{1} &= g_{1}\, \xi^{2}/2,\\ 
\Omega_{2} &= - g_{2}\,\xi-g_{2}\dfrac{2V_{1}}{V_{2}}\displaystyle\int_{0}^{\xi} \cos\left[ \delta\left( \bar{\tau}_{2}-\bar{\tau}_{1}\right) \right]\,d\xi,\\
\bar{\tau}_{2}&=-\displaystyle\int_{0}^{\xi}\Omega_{2}\,d\xi,
\end{array} 
\label{eq3}
\end{equation}
where $ g_{j}=\frac{1}{2}\kappa\pi\delta^{2} \mathrm{csch}\left( \pi\delta/2V_{j}\right)  $. Hence, the first (leading) soliton will always linearly redshift in the frequency domain with rate $ g_{1} $, and decelerate in the time domain. Whereas the dynamics of the second (trailing) soliton depends on two different components: (i) its own (self) component that will lead to a linear redshift similar to the leading soliton, with rate $ g_{2} $; (ii) a cross component representing the effect of the first soliton on the second soliton. The latter component is proportional to the ratio between their amplitudes and the cosine of the time difference between them. Since the cosine term varies between positive and negative values, the dynamics of the second soliton can switch back and forth between redshift and blueshift in the frequency domain or deceleration and acceleration in the time domain. This analytical model shows a very good agreement with the numerical model as depicted in Fig. \ref{Fig1}, provided that the assumption of the soliton durations $ \ll 1/\delta $ is satisfied. Deviations from this condition results in an overestimated value of the frequency-shift rate, $ g $, of an about a factor of 2. In Fig. \ref{Fig1},  $ t=\tau t_{0} $, $ z=\xi z_{0} $ are the physical delay and propagation distance, respectively, $ z_{0}= t_{0}^{2}/\left|\beta_{2} \right|$, $t_{0}$ is the pulse duration, $\beta_{2}$ is the second-order dispersion coefficient. Also, this method can be extended easily to study the interactions of more than two solitons.

\begin{figure}
\centerline{\includegraphics[width=8.6cm]{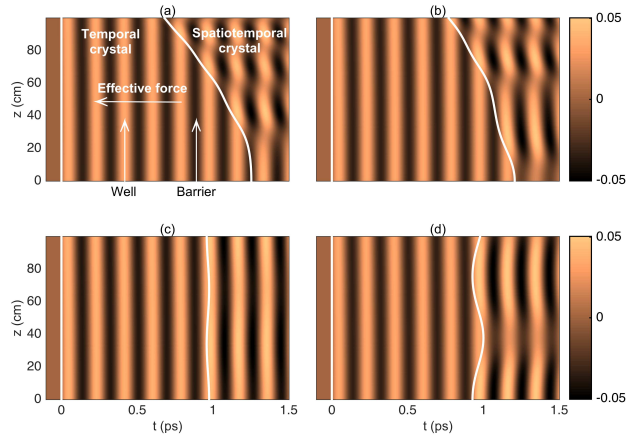}}
\caption{(Color online). Raman polarization induced by two fundamental solitons propagating in the gas-filled HC-PCF, used in Fig. \ref{Fig1}, in a reference frame moving with the leading soliton. White solid lines represent the temporal evolution of  the two solitons using the analytical prediction [Eq. (\ref{eq3})]. The second soliton is launched at (a) $ \bar{\tau}_{2}\left(0 \right) =6.75\bar{\Lambda}$, (b) $ \bar{\tau}_{2}\left(0 \right) =6.5\bar{\Lambda}$, (c) $ \bar{\tau}_{2}\left(0 \right) =5.25\bar{\Lambda}$, and (d) $ \bar{\tau}_{2}\left(0 \right) =5\bar{\Lambda}$. 
\label{Fig2}}
\end{figure}

Figure \ref{Fig2} shows four special cases of the temporal evolution of the second soliton superimposed on the total induced-Raman polarization in a reference frame moving with the leading soliton $ \tilde{\tau}=\tau-g_{1}\xi^{2}/2 $. $ V_{2}$ is chosen smaller than $ V_{1} $ so that the cross component in Eq. \ref{eq3} is comparable to the self component. The leading soliton induces a lagging sinusoidal temporal modulation of the medium refractive index due to the Raman effect, resulting in a {\em temporal crystal} with period $ \bar{\Lambda} $. Due to soliton acceleration or spectral redshift, a constant force $ -g_{1} $ acts on the crystal in the positive-delay (right) direction \cite{Saleh15a}. Hence in the accelerated moving frame, there is an effective force acting on the trailing soliton on the other direction, similar to a particle in a moving periodic potential. The uniformity of the crystal is modified along the direction of propagation, resulting in a spatiotemporal modulation of the refractive index, i.e. a {\em spatiotemporal crystal} \cite{Biancalana07}. As we are operating in the anomalous dispersion regime, the positive (negative) variation of the refractive index represents a potential well (barrier). Based on the initial time delay between the two solitons $ \Delta\bar{\tau}_{i} $, the dynamics of the second soliton  is different, and the shape of the crystal is altered. Looking at Fig. \ref{Fig2}, launching the second soliton at (a) the top of a barrier or (b) at the right edge of a well induced by the first soliton, the second soliton will be able to overcome the barriers during propagation, and transported across the potential by the applied force to the left direction. The output spatiotemporal crystals are chirped along the direction of propagation in these cases. Interestingly, the second soliton in (b) experiences a net maximum self-frequency blueshift of 9.12 THz $ \equiv 51.4 $ nm after 10 cm of propagation, before it is redshifted. Launching the second soliton at (c) the minimum or (d) a left edge of a well, the second soliton will not be able to overcome the barriers in these cases, so it is trapped inside the well and will oscillate indefinitely. The amplitude of oscillation in (c) is very small, since the initial velocity of the soliton in this potential is zero. The soliton will oscillate in an asymmetric manner due to the modified potential beyond the second soliton peak as well as the acting force that is opposite to the initial velocity as in (d). The resulting spatiotemporal crystals have uniform periods along the direction of propagation in both cases.

\begin{figure}
\centerline{\includegraphics[width=8.6cm]{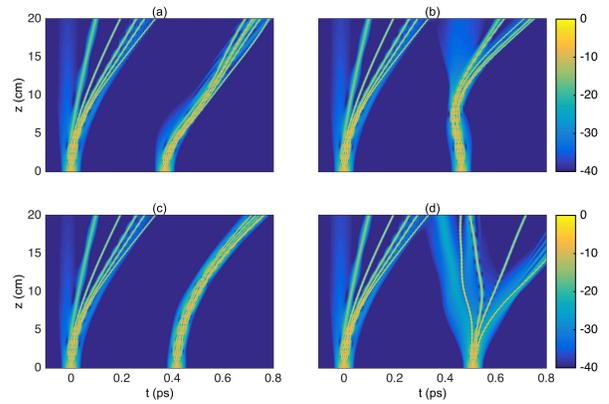}}
\caption{(Color online). Temporal evolution of two successive identical Gaussian pulses with profile $ N\exp\left(-\tau^{2}/2 \right)$ in the gas-filled HC PCF, with $ N=7 $, and FWHM = 25 fs. The first pulse is launched at $ \bar{\tau}_{1}\left(0 \right) =0$. The second pulse is launched at: (a) $ \bar{\tau}_{2}\left(0 \right) =2\bar{\Lambda}$, (b) $ \bar{\tau}_{2}\left(0 \right) =2.5\bar{\Lambda} $, (c) $\bar{\tau}_{2}\left(0 \right) =2.25\bar{\Lambda}$, and (d) $\bar{\tau}_{2}\left(0 \right) =2.75\bar{\Lambda} $.
\label{Fig3}}
\end{figure}

The temporal evolution of two successive ultrashort Gaussian pulses rather than fundamental solitons are portrayed in Fig. \ref{Fig3} for different time delays. The two pulses have the same central frequencies and amplitudes, and the delay is within the relaxation coherence time of the Raman-active gas. The two pulses will experience pulse compression and soliton fission processes. The leading `first' pulse excites the Raman polarization `potential' that will affect the trailing `second' pulse. The dynamics of the leading pulse is independent of the time delay and will encounter a self-induced Raman redshift (deceleration). The trailing pulse dynamics is influenced by its Raman-induced effect as well as the cross-accelerated Raman polarization effect of the leading pulse. In Fig. \ref{Fig3}, panel (a) shows the case when the self and cross components are working together, resulting in a strong delay in comparison to the first pulse. The situation when the self and cross components acts against each other is figured in panel (b), where initially the second pulse is nearly halted since the cross and self components cancel each other. When the second pulse is launched at one minima of the potential induced by the first pulse, the pulse is well-confined during propagation, even after the pulse fission the generated solitons are still traveling together, see panel (c). Launching at one potential-maxima, the dynamics of a tree-like behavior is obtained as shown in panel (d), where each soliton propagates in a different direction. In all the cases, the dynamics of each soliton depends on where exactly this soliton is emitted inside the total accelerated periodic potential induced by other preceding solitons.

In conclusion, the propagation of two successive ultrashort strong pulses in HC-PCFs filled by Raman-active gases have been investigated. The two pulses are not temporally overlapped and separated by a delay smaller than the Raman polarization dephasing time ($\sim100$ ps). We have adopted  perturbation theory to understand the effect of the Raman nonlinear polarization on the dynamics of the two pulses, assuming that they are fundamental solitons with durations much less than the Raman transition period. We have found that the leading soliton will always redshift, while the trailing soliton dynamics can oscillate between redshift and blueshift processes based on the time delay between the two solitons. The second soliton can experience a boosted blueshift without experiencing losses, in contrast to the ionization-induced blueshift technique \cite{Hoelzer11b,Saleh11a}. The leading soliton induces an accelerated temporal crystal due to the long-living Raman polarization. This crystal is upgraded to a spatiotemporal one with a uniform or chirped spatial period beyond the trailing soliton. Our simulations have also been extended to include the case of two energetic Gaussian pulses. We have found that a good control of the trailing pulse dynamics can be achieved via changing the time delay between the two pulses. These theoretical results  will  pave the way to manipulate and control pulse dynamics in PCFs for novel optical applications.

We would like to acknowledge several useful discussions with Prof. Philip St.J. Russell,  Dr. John C. Travers, and Dr. Amir Abdolvand in the Max Planck Institute for the Science of Light in Erlangen. The authors would like also to acknowledge the support of this research by the Royal Society of Edinburgh,  Scottish Government, and the German Max Planck Society for the Advancement of Science.

\bibliographystyle{osajnl}	

\end{document}